\documentclass[twocolumn,showpacs,preprintnumbers,superscriptaddress,pre]{revtex4}
\usepackage{amsfonts,amssymb,amsmath,latexsym,epsfig} 
\usepackage{graphicx}
\usepackage{amssymb}
\usepackage{epstopdf}
\usepackage{natbib}
\usepackage{hyperref}
\usepackage{verbatim}
\usepackage{commath}
\usepackage{color}
\definecolor{greencolor}{rgb}{0,0.5,0.2}
\definecolor{redcolor}{rgb}{.7,0.,0.}
\definecolor{bluecolor}{rgb}{0,0.,1.}
\definecolor{greycolor}{rgb}{.5,.5,.5}

\begin{document}
\title{Temporal-varying failures of nodes in networks}
\author{Georgie Knight}
 \email{georgiesamuel.knight@unibo.it}
\affiliation{%
Dip. Matematica, Universit\`a di Bologna, Piazza di Porta San Donato 5, 40126 Bologna, Italy
}%
\affiliation{%
Institute of Mathematics, The Hebrew University, Jerusalem 91904, Israel
}%
\author{Giampaolo Cristadoro}
 \email{giampaolo.cristadoro@unibo.it}
\affiliation{%
Dip. Matematica, Universit\`a di Bologna, Piazza di Porta San Donato 5, 40126 Bologna, Italy
}%
 \author{Eduardo G. Altmann }
\email{edugalt@pks.mpg.de}
\affiliation{%
Max Planck Institute for the Physics of Complex Systems, Dresden,Germany
}%

\date{\today}
\begin{abstract}
We consider networks in which random walkers are removed because of the failure of
specific nodes. We interpret the rate of loss as a measure of the importance of nodes, a notion we denote as
\emph{failure-centrality}.  We show that the degree of the node is not sufficient
to determine this measure and that, in a first approximation, the shortest loops through
the node have to
be taken into account. We propose approximations of the failure-centrality which are
valid for temporal-varying failures and we dwell 
on  the possibility of  externally changing the relative  importance of nodes in a given
network, by exploiting the  interference  between the loops of a node and the  cycles of
the temporal pattern of failures.  In the limit of long failure cycles we show
  analytically that the escape in a node is larger than the one estimated from a
  stochastic failure with the same failure probability. We test our general formalism in
 two real-world networks   
  (air-transportation and e-mail users) and show  how communities lead to  deviations from
  predictions for failures in hubs.
\end{abstract}
\maketitle

\section{Introduction}

Random walks in networks are the core of many complex-systems models\cite{NewmanBookNetworks}.
 Here we consider the case of \emph{open} networks  in which there is a positive probability of a random-walker
being lost  (or equivalently to freeze its movement) on a node (e.g., because of its \emph{failure}). This situation is also
known as the {\em trapping problem}, which after the pioneering work of Montroll
\cite{Mon69}, has received much recent attention  \cite{TejEtAl09,LinZhang13} \footnote{This is not to be confused with
the process of key-removal, where nodes and edges are removed but the network remains closed and no trapping mechanism is in action (see for example \cite{New08}, \cite{ResEtal06})}. Its relevance  is apparent, for instance, in communication networks where a failure could correspond
to the loss  of an information package. Despite the negative connotation of the term {\it
  failure}, the same picture describes also situations in which the removal of random
walkers is deliberately done. For instance, the removal of the information package at a
node could reflect the successful arrival at the target, similar to search-related problems  \cite{AdEtAl01,GuiEtAl02,Holme03}.  These ideas are not restricted to
communication networks, e.g. in epidemic modeling failure of specific nodes could
correspond to infected locations or the loss of a virus due to immunisation.  

\begin{figure}[hbt!]
\begin{center}
\hspace{0.1cm} \includegraphics[width=4cm]{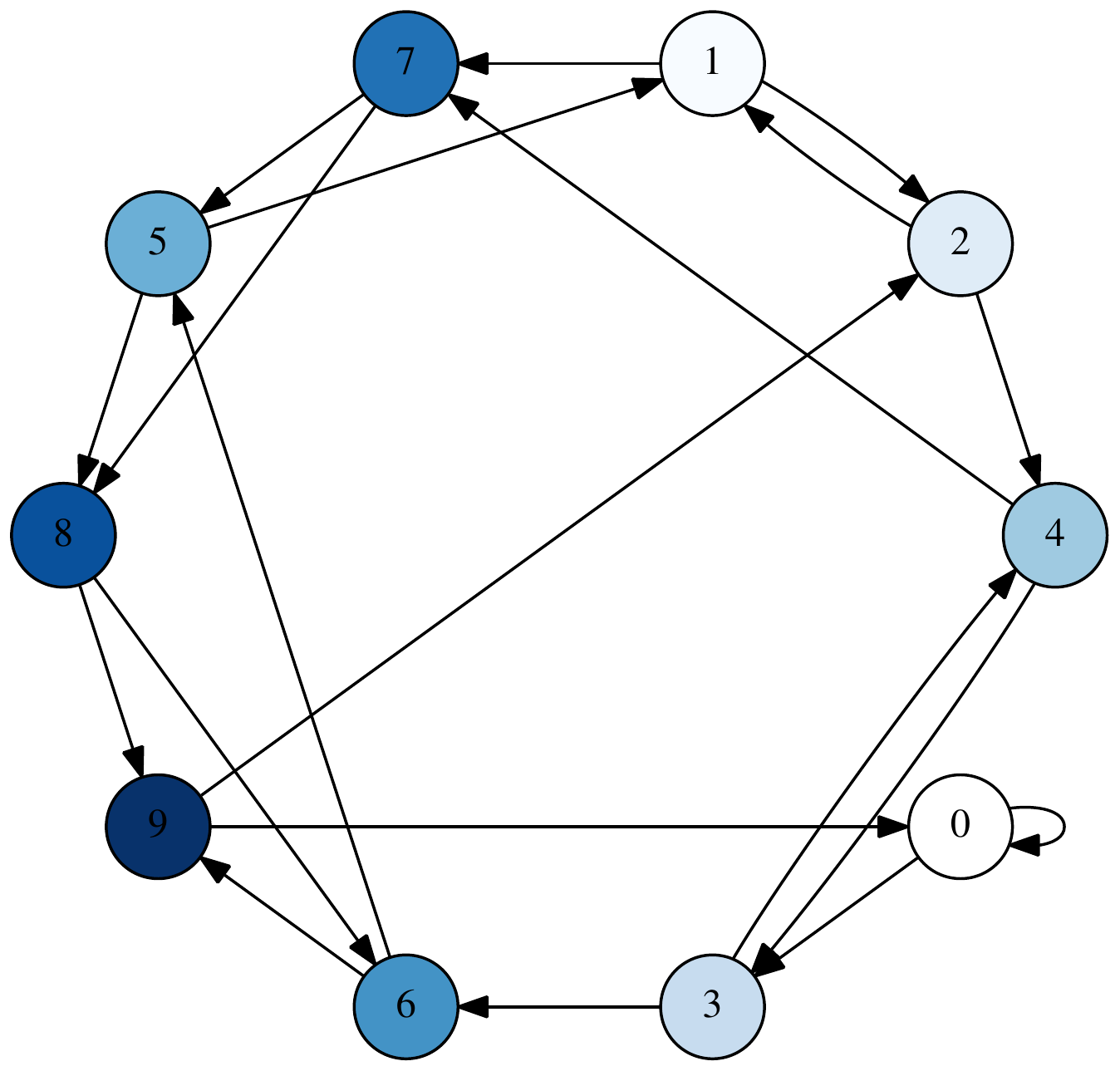}\hspace{0.3cm}  \includegraphics[width=4cm]{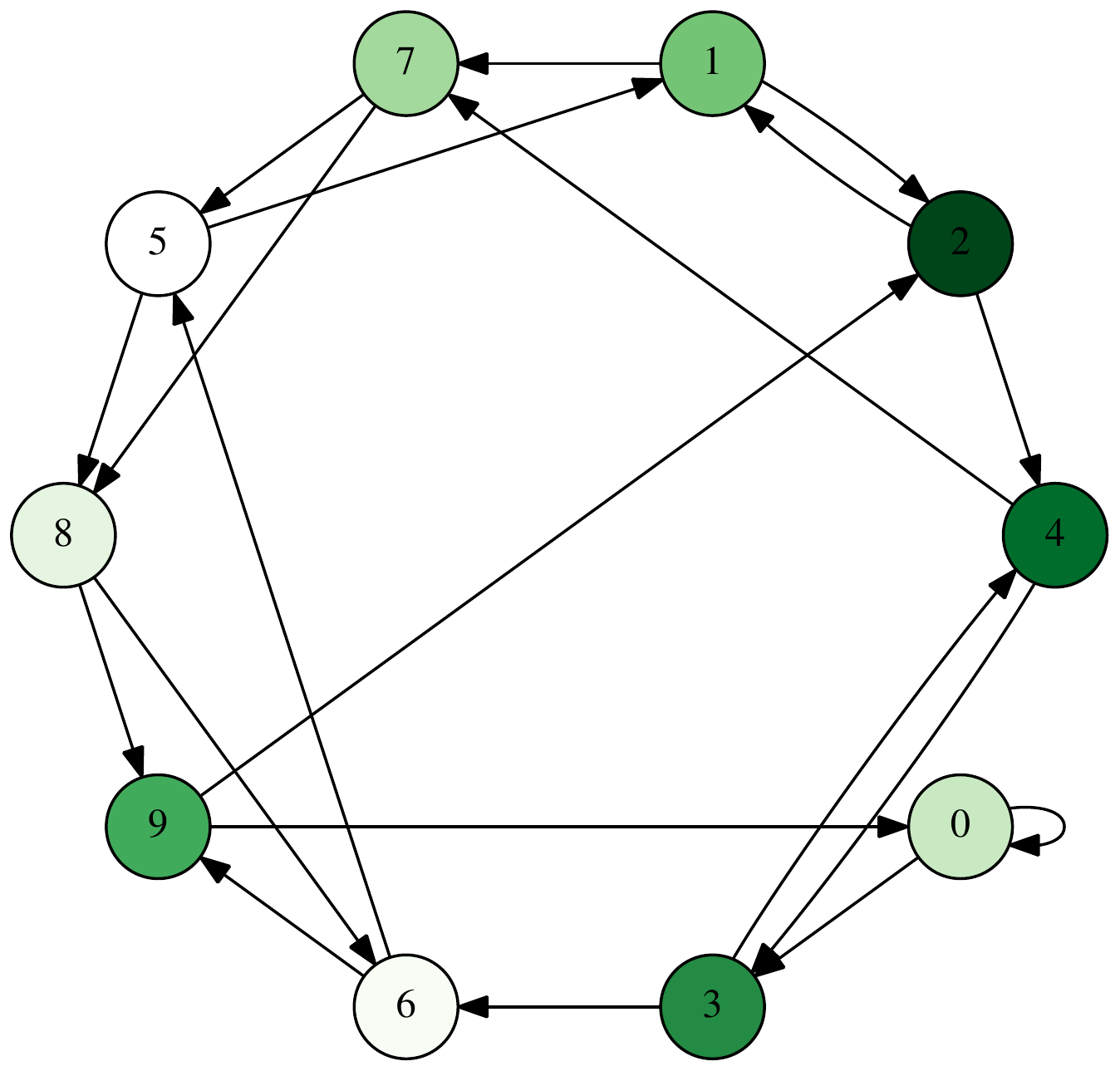}\\
 \hspace{-0.3cm}\includegraphics[width=4.2cm]{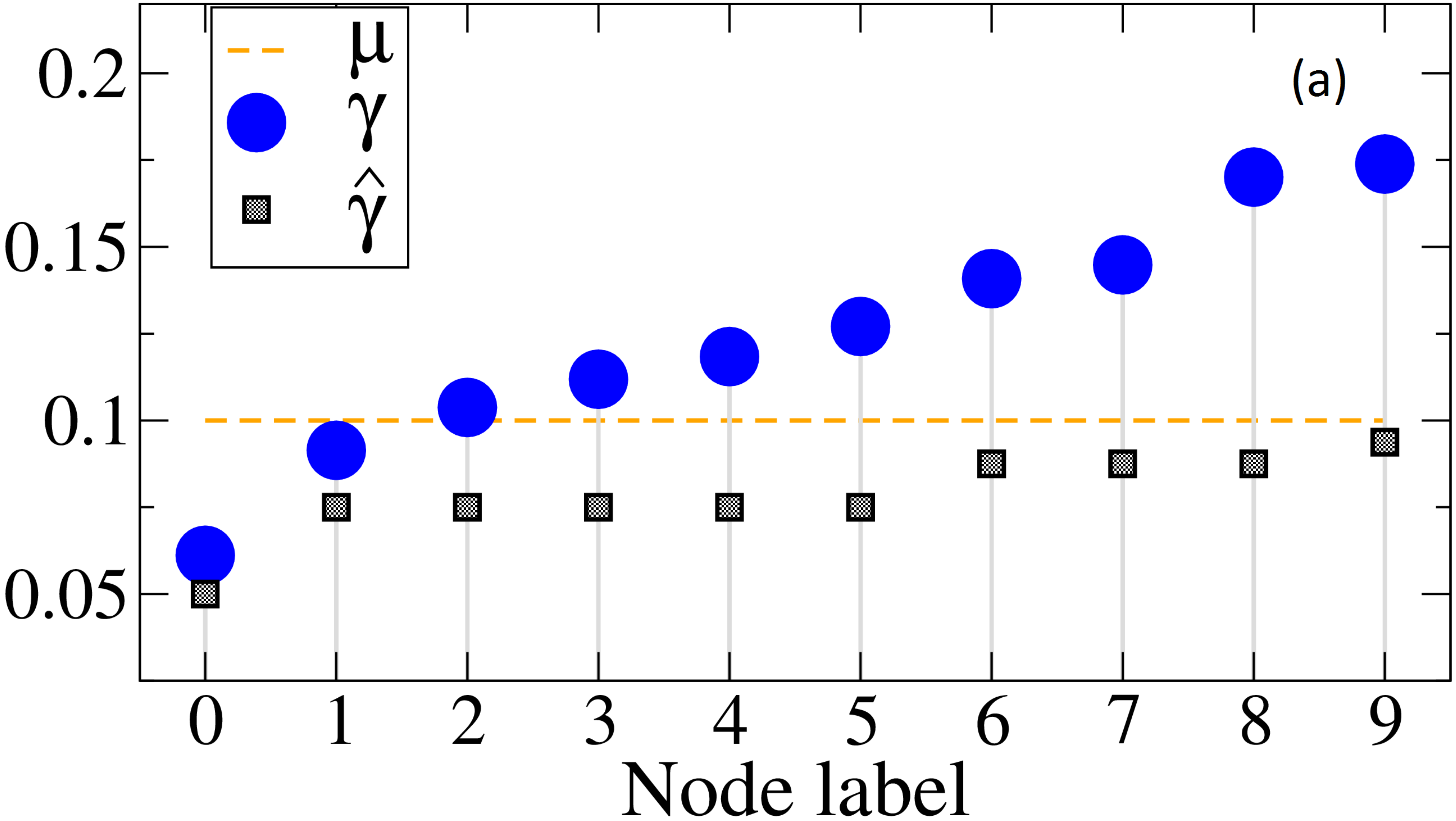} \hspace{0.15cm} \includegraphics[width=4.2cm]{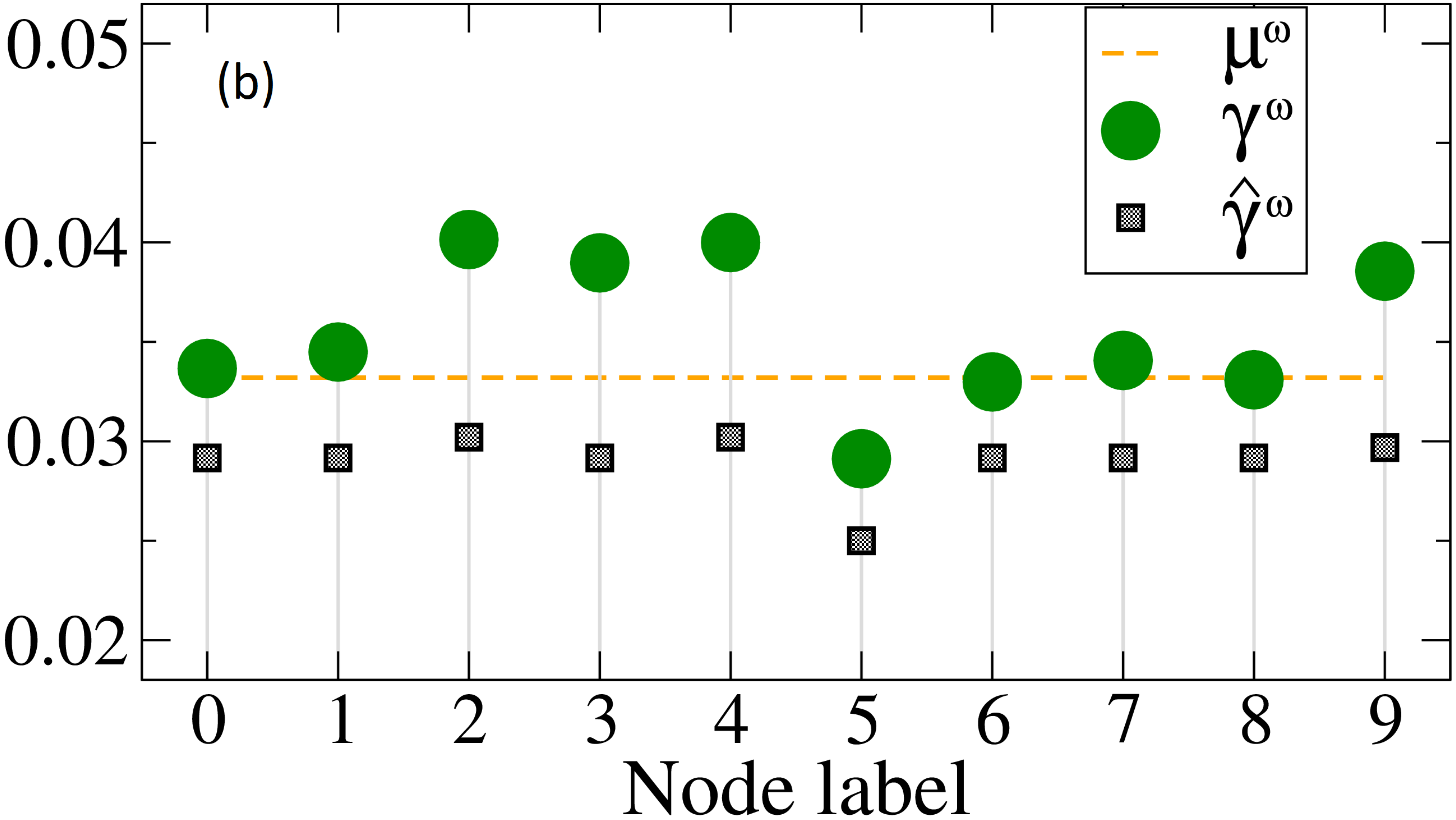}
\end{center}
\caption{(Colour online) \emph{Node dependency of the escape rate~$\gamma$}. 
The ten nodes of a directed regular network with (in- and out-) degree $2$ are
labeled according to the ranking of the escape rate in the case of complete
failure. Panels (a,b) full circles show ~$\gamma$ obtained when each node fails (a) completely or
(b)  periodically in time, with period three ($\omega=[110]^{\infty}$). Zero-order (dashed
line) and first-order (squares) approximations  are  shown  (see main text). For such small
networks, the first order approximation of the {\it value} is not substantially better,
but the {\it ranking} of nodes agrees much better with the observations.  Nodes are
colored according to $\gamma$.
}
\label{fig:Switching}
\end{figure}

Measures for the importance (or centrality) of nodes and edges are also often
defined based on random walks~\cite{NewmanBookNetworks,Noh04,PerraEtal12,Newman2005}. 
In many situations the importance of a node or link is manifest only after its
\emph{failure}.  This motivates us to consider the rate of the exponential loss from a network  in which a
node fails (escape rate) as a measure of the \emph{failure-centrality} of this node. The traditional analysis
of trapping problems is based on the first passage time to the considered
node~\cite{NewmanBookNetworks,RednerBookFirstPassage}.  A quantity closely related to the
escape rate considered here is the global mean first-passage time, and a typical problem
is how this quantity scales with network size (see
e.g. Refs. \cite{TejEtAl09,LinZhang13}). Mean-field models allow for an
estimation of the mean first passage time (and also the escape rate) for nodes of a given degree (see, e.g.,
Refs.~\cite{Baronchelli2006,Lau2010}).
A point that seems to have not been explored in detail so far is that nodes with the same
degree may show very different behavior, i.e., our failure-centrality differs from the
degree of the node.   An example of this variability  is shown in
Fig.~\ref{fig:Switching}. Nodes leading to a faster decay of the number of surviving walkers are the most 
important one, either because they lead to the strongest leakage of the network or because
they provide the fastest target for random-walkers.

The situation sketched above, equivalent to an  absorbing Markov chain, has a parallel in  the problem of placing holes in the phase space of
chaotic dynamical systems, as previously noticed and explored in Refs.~\cite{Afraimovich2010,Geo12,BunBook,BocEtal12}.  In that context, the dependence of the asymptotic escape rate on the size and
position of the hole has been a topic of much recent
interest~\cite{Bun11,Geo12,Cr13,RMP,Bahsoun,Livorati}. One insightful question
is~\cite{Bun11}: Where to place a hole to achieve a maximal escape rate? In the network
context, this corresponds to asking~\cite{Afraimovich2010}: Which is the best node to place the target on so that
it is faster/slower for a random walker to find it?
Some analogies between open dynamical
systems and networks has been used also in applications of isospectral
transformations that reduce the complexity of  networks~\cite{BunBook} and in analyses of the connectivity of the network~\cite{BocEtal12}.

In this paper we investigate the network properties responsible for the variety in the
escape rate obtained when nodes fail. We adapt previously
known results to the network problem (as in
Refs.~\cite{Afraimovich2010,Geo12,BunBook,BocEtal12}) and we then extend them   to
temporal-varying 
failures. In particular,  we show how the escape rate obtained when one node fails can be approximated
based on the short loops passing through this node, one element which can often be
neglected in random networks but which is present in
virtually all real world examples~\cite{NewmanBookNetworks,Martin2010}. 
 We dwell on the possibility of  externally changing the relative  importance of nodes in a given
network, by exploiting the  interference  between the loops traversing the nodes and
externally-imposed failures. Finally, we test our results in two  real-world networks (air
transportation and e-mail) and show how community structures influence our estimations.

\section{Setting}
The process we study is a random walk on the network  defined  by an $N\times N$  irreducible and aperiodic
stochastic  matrix $\mathbf{A}_1$:
its entries $a_{jk}$ are the  probability for a random walker to jump  from node $j$ to
node $k$ at every (discrete) time step.  
Starting from a weighted, fully-connected, directed network with $N$ nodes -- defined by the
$N\times N$ connectivity matrix $\mathbf{W}$ with real-valued entries  ${w}_{jk}$ --
$\mathbf{A}_1$  may be constructed considering $w_{jk}$ to be proportional to the probability to follow the outgoing link ($a_{jk}=w_{jk}/\sum_kw_{jk}$).
We call the dynamics in $\mathbf{A}_1$ the \emph{closed} dynamics. We can \emph{open} it
by letting a node  \emph{fail}:  the failure of a node $j$ is modeled by modifying its outgoing links ${a}_{jk}$  - eventually as a function of time - in such a way that $\sum_{k} {a}_{jk}<1$. This corresponds to the physical intuition that a walker on the failed node has a non-zero probability to be lost from the network.

In a network with a failed node $j$, the probability $P_n(j)$ that a random walker is not lost up to time $n$   asymptically decays exponentially with a rate
\begin{equation}
                          \gamma(j)= - \lim_{n\rightarrow \infty} \frac{1}{n}\ln (P_n(j)).
\label{Eq:Escape-rate-definition}
\end{equation}
We interpret the escape rate $\gamma(j)$ as the {\it failure-centrality} of the node $j$,
i.e., a measure capturing the sensitivity of the network  to its failure.

Consider first the case of total failure of a node $j$, corresponding to the case where  all its outgoing links  have zero weight (equivalent  to certain loss  of the random walker once  on the node); and denote by $\mathbf{A}_0(j)$ the corresponding matrix. It is well known that
\begin{equation}
                          \gamma(j)=-\ln(\| \mathbf{A}_0(j)\|)
\label{EQ:Escape-rate-eigenvalue}
\end{equation}
where $\| \mathbf{\cdot} \|$ denotes the largest eigenvalue of the matrix. Note that, according to the Perron-Frobenius Theorem, $\| \mathbf{A}_1\|=1$  corresponding to no loss of walkers moving randomly through the network.
Generically $\gamma(j)$ is a non trivial function of the node, even for very simple,
regular networks such as that displayed in Fig.~\ref{fig:Switching}a. 
 In applications, it is important to determine which node
is more relevant (in the above sense) for a given network, and more generally their
ranking in  relative importance. 
In principle, this requires generating the open network
  $\mathbf{A}_0(j)$ and computing  Eq.~(\ref{EQ:Escape-rate-eigenvalue}) for $j=1 \ldots N$ nodes. In practice, for large $N$ this
  requires a prohibitive computational  effort (see~\cite{Cr13} for an analytic approach that works in some particular cases). Furthermore, such brute-force computation
  brings no insight on the properties of  the closed network  $\mathbf{A}_1$ responsible
  for the importance of a node. 
Here we show how the values and ordering of $\gamma(j)$ can be efficiently estimated based on the degree and short loops of node~$j$ in $\mathbf{A}_1$. 

\section{Escape rate approximants}

As a  first  approximation we can imagine that, if the network is complex enough, at each time step it looses a fraction of walkers corresponding to the equilibrium measure~$\mu(j)$  of  the closed network on the failed
node~$j$, so that  asymptotically in time $P_n(j) \sim (1-\mu(j))^n$ and thus for a large network ($\mu(j)<<1$)
  we obtain as the zero-th order approximation 
\begin{equation}\label{eq.gamma0}
\gamma(j) \simeq \mu(j).
\end{equation}
In networks with fixed degree ($k$-regular networks),  Eq.~(\ref{eq.gamma0}) predicts
  the  same $\gamma$ for all nodes~$j$, as does eigenvector centrality and
  PageRank of the closed network~\cite{NewmanBookNetworks}. In chaotic systems, equivalent approximations have been shown to provide only a rough
estimate of $\gamma(j)$ which, even in the case of uniform stationary distribution, show
interesting fluctuations (see Ref.~\cite{RMP} for references and an historical overview, and Refs.~\cite{KeLi09, Bun11}, for recent rigorous connections between the fluctuations and {\em periodic
  orbits}).  
Adapting these results to the problem of random-walk in complex networks, we obtain that a better approximation $\hat{\gamma}$ is 
\begin{equation}
                              \hat{\gamma}(j)= \mu(j)\left(1-(\mathbf{A}^{n_j})_{jj}\right)
\label{Eq:Gamma-first-order-form}
\end{equation}
where $n_j$ is the length of the shortest loop through the node $j$ and 
$(\mathbf{A}^{n_j})_{jj}$ is the probability for a walker  to return to the node $j$   after $n_j$ steps, through its shortest loops.
 An example of the approximation~(\ref{Eq:Gamma-first-order-form}) is  illustrated in Fig.~\ref{fig:Switching}a together with the zero-  and first-order approximations.

In the following, we consider the case of temporal dependent failures.  In chaotic dynamical
systems,  holes having a deterministic~\cite{Livorati} and stochastic~\cite{Bahsoun}
variation in the {\it position} have been considered.  Here we fix the failing node and
consider a (stochastic and deterministic) temporal variation in its state (failing or not).

\subsection{Temporal-varying failures - fixed protocol} 

Consider  the situation in which a given temporal fixed protocol of failures, defined by a periodic binary sequence $\{\omega_t\} :=\omega_1 \omega_2  \cdots \omega_p \omega_1 \omega_2 \cdot \omega_p\cdots=[\omega_1\cdots\omega_p]^{\infty}$   is externally imposed on a given node $j$. More precisely,  at each time $t$  a random walker on the network moves in the closed network $\mathbf{A}_1$ if  $\omega_{t}= 1$  or in the open  $\mathbf{A}_0(j)$ otherwise.  In such a situation, the probability for the walker not to be lost after $n$ iterations 
decays exponentially with a rate
 $\gamma^{\omega}(j)$  derived from the corresponding product of the matrices  
\begin{equation}
                            \gamma^{\omega}(j) =-\frac{1}{p}\ln \left( || \prod_{t=1}^p \mathbf{A_{\omega_t}}||\right),
\label{Eq:Definition_FER}
\end{equation}
where, to lighten notation, we avoid explicitly expressing the dependence on the node and simply denote $A_0$ the open network.

We now derive approximations of Eq.~(\ref{Eq:Definition_FER})  in the same spirit as the
approximations to Eq.~(\ref{EQ:Escape-rate-eigenvalue}) described above. The zero-th
order approximation analogous to Eq.~(\ref{eq.gamma0}) is obtained re-scaling the equilibrium measure  by the
fraction of time $r_{\omega}:=\sum_{t=1}^p(1-\omega_t)/p$ the node $j$ fails over a period
$p$ as 
\begin{equation}\label{eq.gamma0t}
\gamma(j) \simeq \mu(j)/r_w.
\end{equation}
A correction term should take  into account loops traversing the failed node, as in Eq.~(\ref{Eq:Gamma-first-order-form}). Note however that in a 
temporal varying  situation it is possible that, for a given protocol $\omega$, starting with  a failure event, the node $j$ is not failing again after a complete traversal of its shortest loop.  Indeed, let $L_j$ be the {\em set} of  lengths of all primitive loops (i.e. that are not a repetition of smaller loops)  of the node $j$. 
For each time $t\le p$ such that $\omega_t=0$ in a given protocol $\omega$, consider the
minimum time $s$ such that $s= k l$ for some $k\in \mathbb{N}^+$ and $l \in L_j$ and
such that  $\omega_{t+s}=0$. 
This time $s$ corresponds to the length of a loop connecting the node $j$ to itself in
times in which it fails and is thus a generalization to the shortest loop in Eq.~(\ref{Eq:Gamma-first-order-form}).
We denote by $T^{\omega}(j)$ the set of all such times  $s$
for the node $j$ under the failing protocol $\omega$. Note that the cardinality of $T$, $|T|$, is equal to the number of failures in a period of $\omega$.
 For example, consider a node  with $L=\{2,5 \}$. For a protocol
 $\omega=[011]^{\infty}$ then $T=\{6\}$, while for $\omega=[01011]^{\infty}$ then $T =
 \{2,5\}$.
Finally, for every $s \in T^{\omega}(j)$, $(\mathbf{A}^s)_{jj}$ denotes the probability  for the walker to return
to the node $j$ in $s$ steps.
An heuristic formula  approximating $\gamma^{\omega}(j)$ considers the average over such probabilities as follows:
\begin{equation}
                          \hat{\gamma}^{\omega}(j)= \frac{\mu(j)}{r_{\omega}}\left(1- \frac{1}{|T |}\sum_{s \in T}{(\mathbf{A}^s)_{jj}}\right).  
                          \label{Eq:gamma_ap}
\end{equation}
The case of trivial protocol $\omega=[0]^{\infty}$,  $|T|=1$, $s$ equals the shortest
loop, and Eqs.~(\ref{Eq:gamma_ap}) and~(\ref{Eq:Gamma-first-order-form})
coincide. For non-trivial protocols, the correction term in Eq.(\ref{Eq:gamma_ap}) accounts for the loops in the network that traverse the blinking node $j$ at times corresponding to failure events.

We tested Eq.(\ref{Eq:gamma_ap}) in two standard ensembles of random networks: undirected with power-law degree
distribution (scale free); and directed k-regular (regular), i.e., all nodes with the same in and
out-degree).
Firstly, we assess the quality of Eq.~(\ref{Eq:gamma_ap}) -- which is based only on local
properties -- as an approximation of the global escape rate.
The percentage error of the naive approximation based on $\mu$ -- Eqs.~(\ref{eq.gamma0}) and
(\ref{eq.gamma0t}) -- and of our approximation are shown in Table \ref{Table:AppAcc}. 
Scale-free networks show a large inaccuracy in the naive approximations, which is greatly
improved with our proposed formula. For the regular
networks, the naive approximation already provides a quite accurate approximation, yet
Eq.(\ref{Eq:gamma_ap}) still represents an improvement.  Both approximations
overestimate $\gamma$ in the scale free network and underestimate it in the regular
  networks (see Fig.~\ref{fig:Switching}). The important factor determining this
  difference is not the degree distribution, but  the directionality of the links (present
  in the  regular but not in the scale free case). Random walkers can travel back and forth
  through undirected links and therefore undirected networks have a higher number of
  short loops (e.g., every link creates a period 2 loop), which tend to reduce $\gamma$
  [as in   Eq.~(\ref{Eq:Gamma-first-order-form})]. The underestimation observed in the directed
  regular network agrees with what is known for the complete failure case in
  chaotic systems~\cite{Geo12}). 
\begin{table}
\begin{tabular}{|l|cc|cc|}
\hline
Error Between $\rightarrow$ & $ (\gamma, \mu)$ &$ (\gamma,\hat{\gamma})$     & $(\gamma^{\omega}, \mu^{\omega})$& $(\gamma^{\omega},\hat{\gamma}^{\omega})$ \\
Network  Topology $\downarrow$&    \multicolumn{2}{c|}{(complete failure)}   &  \multicolumn{2}{c|}{(temporal failure)}\\ 
\hline
Scale free  &   $33.6 (4)$     & $5.5(1)$  & $13.0(1)$  &    $1.24(3)$ \\
 Regular  & $0.695(6)$ &  $0.650(1)$  & $0.336(2)$   & $0.3301(4)$ \\
\hline
\end{tabular}
\caption{ (Colour online) {\em Approximation accuracy}. Percentage error between the approximations
  ($\mu,\hat{\gamma}$) and true ($\gamma$) escape rate computed as an average over all nodes
  in the network. 
  Each node is considered to fail at all times (complete failure) or periodically in time
  with protocol $\omega=[01011]^{\infty}$ (temporal failure).  Random networks with $1,000$ nodes and two different degree
  distributions $P(deg(i)=k)$  were used: scale-free networks are undirected and have
  $P(deg(i)=k)\sim k^{-2.5}$ while 
  regular networks have out-degree equal to in-degree  equal to $3$.  The reported values are the
  average and the standard error of the mean (in brackets, corresponding to an uncertainty
  in the last  digit) computed over an ensemble of networks ($94$ generated
  according to Ref.~\cite{Hag08} for scale free and  $100$ generated according to
  Ref.~\cite{Kim12} for regular networks.)}
\label{Table:AppAcc}
\end{table}

\subsection{Switching the relative importance of nodes} 

The relative importance of a node is not an intrinsic property of the network as it can
depend upon the type of node failure. The most interesting consequence of this is that
the rank order of two nodes may be switched if the failure becomes time dependent
(compare panels $(a)$ and $(b)$ of Fig.~\ref{fig:Switching}).
 To elucidate the
basic mechanism underlying such switching phenomena, consider  the simple  case in which all
nodes have the same equilibrium measure $\mu$, as in the $k$-regular network discussed
above. Suppose for definiteness that two nodes $i,j$ have rates
  $\hat{\gamma}(i) > \hat{\gamma}(j)$ and shortest loops $n(i) <  n(j)$, as expected from Eq.~\ref{Eq:Gamma-first-order-form}). In this  situation, from Eq.~(\ref{Eq:gamma_ap}) we see that a periodic protocol
  of  period  $p=n(j)$ and with a single failure over the period $p$,  will induce a
  switching on the relative importance of the node, that is  $\hat{\gamma}^{\omega}(j) >
  \hat{\gamma}^{\omega}(i)$. 
Here we are using the fact that the importance of a node depends on both its invariant
measure and  first return-time probability, see Eq.~(\ref{Eq:Gamma-first-order-form}).
As an example,  Fig.~\ref{fig:Switching}(b) shows how the importance of nodes $2$
  $(n_2=2)$  and $7$  $(n_7=3)$ are switched by the  protocol $\omega=[110]^{\infty}$.
By exploiting the  interference  between the loops of a node and the  periods of the
temporal pattern of failures it is  thus possible to devise protocols $\omega$ that change
the relative  importance of nodes in a given network.

We now quantify the extent to which Eq.~(\ref{Eq:gamma_ap}) describes switches. For each
node $j$ of each of the two networks discussed in Tab.~\ref{Table:AppAcc} we compute
$\gamma^{\omega}(j)$ and $\hat{\gamma}^{\omega}(j)$ for the protocol
$\omega=[01011]^{\infty}$  and $\gamma(j)$ for the fully open case (i.e., protocol
$[0]^\infty$). This led to three different rankings of the nodes for each network. We first confirm that
the ranking of nodes estimated from $\hat{\gamma}^\omega$ is very close to the true one
obtained from $\gamma^\omega$. This is quantified by the Kendall rank
correlation $\tau$\footnote{ The {\em Kendall's} $\tau$ rank correlation coefficient between two ranked data
  sets ($x_1,x_2,\ldots, x_n$ and $y_1,y_2, \ldots, y_n$) is defined as
  $\tau=\frac{n_c-n_d}{\frac{1}{2}n(n-1)}$, where $n_c$ is the number of concordant pairs
  and $n_d$ the number of {\em discordant} pairs.  (The total number of
  pairs is $n(n-1)/2=n_d+n_c$ ). A pair of elements $(i,j)$ is concordant if  their order in both
  ranking agree (i.e, $x_i \lessgtr x_j \Rightarrow y_i \lessgtr y_j$), otherwise it is
  discordant.} between these two rankings, that yields
$\tau=0.99920(5)$ (scale free network) and $\tau=0.926(1)$ (regular network), where the
distance to $\tau=1$ is proportional to the number of switches needed to map the estimated
into the true ranking. Next we investigate the magnitude of switchings
introduced by the time-periodic protocol $\omega$ and the extent to which our
estimation~(\ref{Eq:gamma_ap}) is able to describe them. We obtain that the protocol leads
to a substantial change in the ranking in the regular network and small but still detectable
change in the scale free network . This is quantified by the {\em Spearman's}  rank 
correlation coefficient $0 \le \rho \le 1$\footnote{The {\em Spearman's} $\rho$ rank correlation coefficient between two ranked data
  sets of size $n$ is defined as 
$\rho=1-\frac{6\sum_i d_i^2}{n(n^2-1)}$,  where $d_i$ is the difference in the ranking
between the $i^{th}$ variables. } between the ranking obtained in the fully ($\gamma$)
and temporal ($\gamma^\omega$) networks: $\rho=0.9640(5)$ (scale free) and
$\rho=0.548(3)$ (regular), where $\rho=1$ corresponds to perfect agreement between the
rankings. The ranking obtained from~$\hat{\gamma}^\omega$ in
Eq.~(\ref{Eq:gamma_ap}) yields a good estimation of these values: $\rho=0.9501(8)$ (scale free) and
$\rho=0.531(3)$ (regular) ($\rho$ computed between the ranks obtained from $\hat{\gamma}^\omega$ and
$\gamma$).

\subsection{Temporal-varying failure - stochastic}

The limit of large periods $p$  in the  periodic-varying scenario studied above can
be thought of as a typical 
realization of a random failure.  Considering that the limit $p\to \infty$ is taken
preserving (on average) the fraction of times $q$ the node fails, the corresponding stochastic process is
that of a node that {\em blinks} on and off at each time step randomly, such that at each
time it is open with probability $q$. 
 Let $\mathbf{G_n}$ be an n-fold product of $\mathbf{A}_0$  and $\mathbf{A}_1$ where the
matrices are chosen randomly with probability $(q-1)$  and
$q$ respectively, and let $\langle  \cdot \rangle$ be the average over such products.
The rate $\gamma^R(j)$ is then defined as the average  rate taken over all possible
$\mathbf{G_n}$ in the limit as $n$ goes to infinity and is given by
\begin{equation}
                           \gamma^R(j) =  \lim_{n \rightarrow \infty}  - \frac{ \langle \ln{ || \mathbf{G_n} ||} \rangle}{n}.
\label{Eq:Definition_RER}
\end{equation}
A picture similar to such a (quenched) case is the (annealed) one in which at every time 
walkers on the failing node  are lost with probability $q$. 
Equivalently, this corresponds to reducing by a factor $1-q$  the 
probabilities to leave the failed node $j$, that is $a_{jk} \to (1-q)a_{jk}$ for all $k$. This
situation can be thought of as the failing node having a degree $q$ of
\emph{permeability}. In this case the rate is 
\begin{equation}
                           \gamma^{P}(j) = -\ln{( ||(1-q) \mathbf{A}_1 +q \mathbf{A_0}(j)|| )}.
\label{Eq:Definition_PER}
\end{equation}
The stochasticity in the two scenarios leading to
Eqs.~(\ref{Eq:Definition_RER}) and~(\ref{Eq:Definition_PER}) differ and satisfy a simple
inequality~\footnote{To see this, notice that: (i) the term inside the logarithmic function in
  Eq.~(\ref{Eq:Definition_PER}) is for $n\rightarrow \infty$ equal to $ || ((1-q)
  \mathbf{A_1} +q \mathbf{A_0})^n ||^{\frac{1}{n}} =  \langle || \mathbf{G_n}  ||
  \rangle^{\frac{1}{n}}$; and (ii) the negative logarithm  is a convex   function and thus (from   Jensen inequality) $-\ln \langle || \mathbf{G_n}  ||  \rangle
  \ge \langle -\ln || \mathbf{G_n}  || \rangle $.}
\begin{equation}
\gamma^{R}(j) \geq \gamma^{P}(j) \quad \quad \forall j,
\label{inequality}
\end{equation}
see  Ref.~\cite{Bodai2013} for an equivalent result in chaotic systems and
Refs.~\cite{Main92,VulBook93} for results on products of random matrices. 
From the point of view of periodic protocols, this implies that for long periods $p$ the
escape rate of most nodes will be larger than the one from the averaged operator.

\section{Real-world networks}

We now test our theory in two networks with topologies created by real interactions. Firstly we study the transport network of the $500$ busiest commercial airports in the
United States in the year $2002$~\cite{Colizza2007}\footnote{ The data and further details
  are accessible at (retrieved Nov. 2014)
  \url{https://sites.google.com/site/cxnets/usairtransportationnetwork}}. Airports (nodes)
are connected (edges) with a strength $w_{i,j}=w_{j,i}$ proportional to the (yearly) number of existing
seats in flights between them. The failure of a node corresponds to an airport to
which aeroplanes arrive but are unable to leave. We then study the communication network of e-mail users at the University Rovira i Virgili \cite{Gu03}\footnote{ The data and further details
  are accessible at (retrieved May. 2015)
  \url{http://konect.uni-koblenz.de/networks/arenas-email}}. The nodes here represent individual e-mail users. An undirected edge exists between users who have exchanged at least one e-mail. Our main motivation for studying these networks
is to test our theory in real-world networks with non-trivial topologies.

\begin{figure}[ht!]
\begin{center}
 \includegraphics[width=8cm]{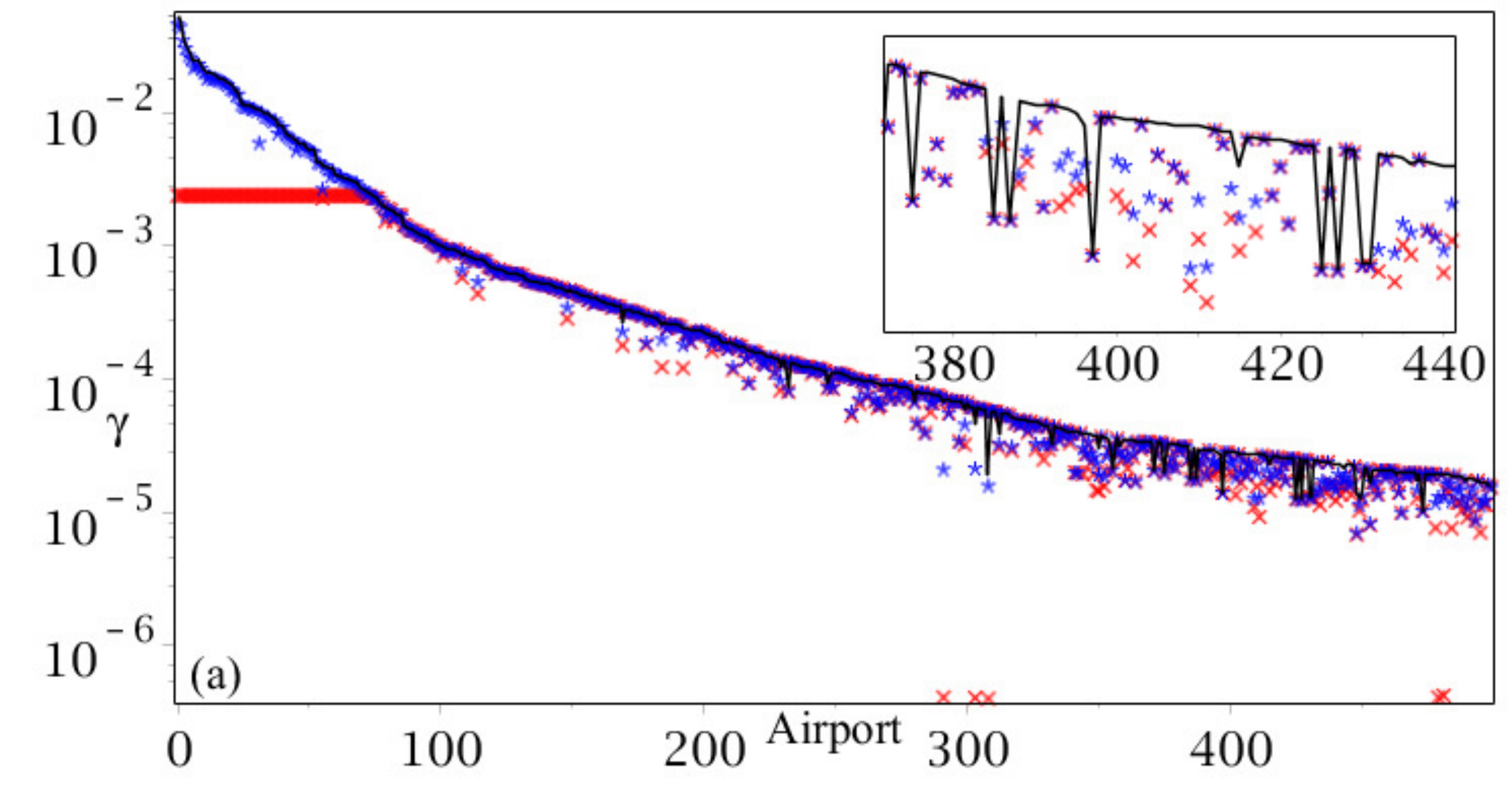}\\
 \includegraphics[width=8cm]{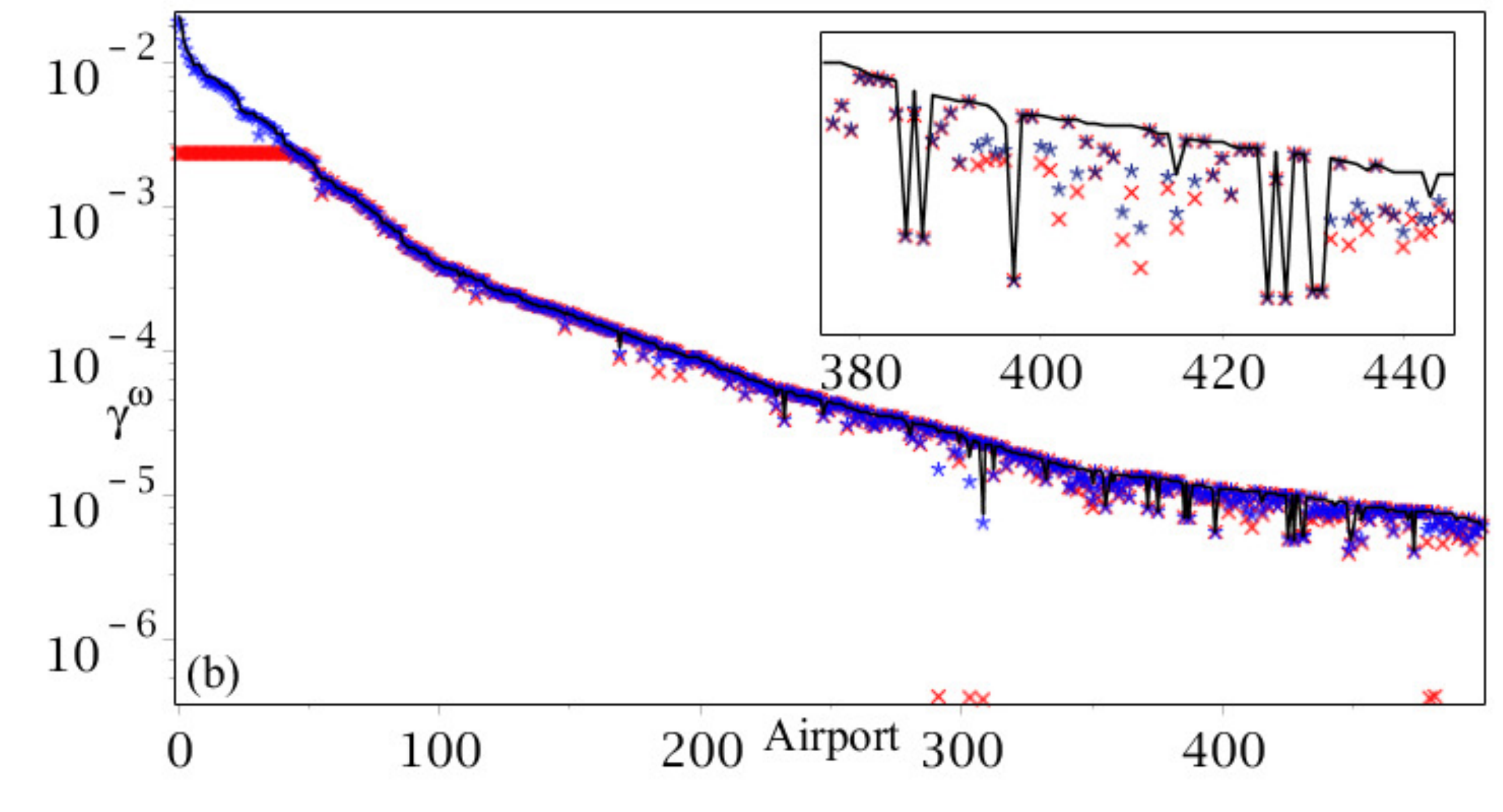}
\end{center}
\caption{(Colour online). Escape rate in the air-transportation network. The plots show the escape rate $\gamma^\omega$ (red
  crosses), the equilibrium measure $\mu$ as in Eq.~(\ref{Eq:Gamma-first-order-form})
  (black line), and the approximation~$\hat{\gamma}$ from Eq.~(\ref{Eq:gamma_ap})
  (blue asterisk) obtained if each of the airports fails. (a) Fully open network
  $\omega=[0]^{\infty}$. (b) Temporal-varying failures with protocol $\omega=[01011]^{\infty}$. The error -- as in Tab.~\ref{Table:AppAcc} -- between $(\gamma, \mu)$
  is $160$ and between $(\gamma,\hat{\gamma})$  is $98.6$. The {\em
    Spearman's}  rank  correlation coefficient $\rho_{A,B}$ between the ranks obtained
  from measures $A$ and $B$ are: 
$\rho_{\text{degree},\gamma^\omega}=$ 0.765,  $\rho_{\mu,\gamma^\omega}= 0.984$, and  $\rho_{\hat{\gamma},\gamma^\omega}=0.997$  for $\omega=[0]^\infty$;
 and
$\rho_{\text{degree},\gamma^\omega}=$ 0.781,  $\rho_{\mu,\gamma^\omega}= 0.993$, and  $\rho_{\hat{\gamma},\gamma^\omega}=0.997$ for $\omega=[01011]^{\infty}$. Finally,
comparing $\gamma^\omega$ of the two protocols we obtain $\rho=0.997$. Values reported to 3 s.f.
}
\label{fig:Airports1}
\end{figure}

\begin{figure}[ht!]
\begin{center}
 \includegraphics[width=8.5cm]{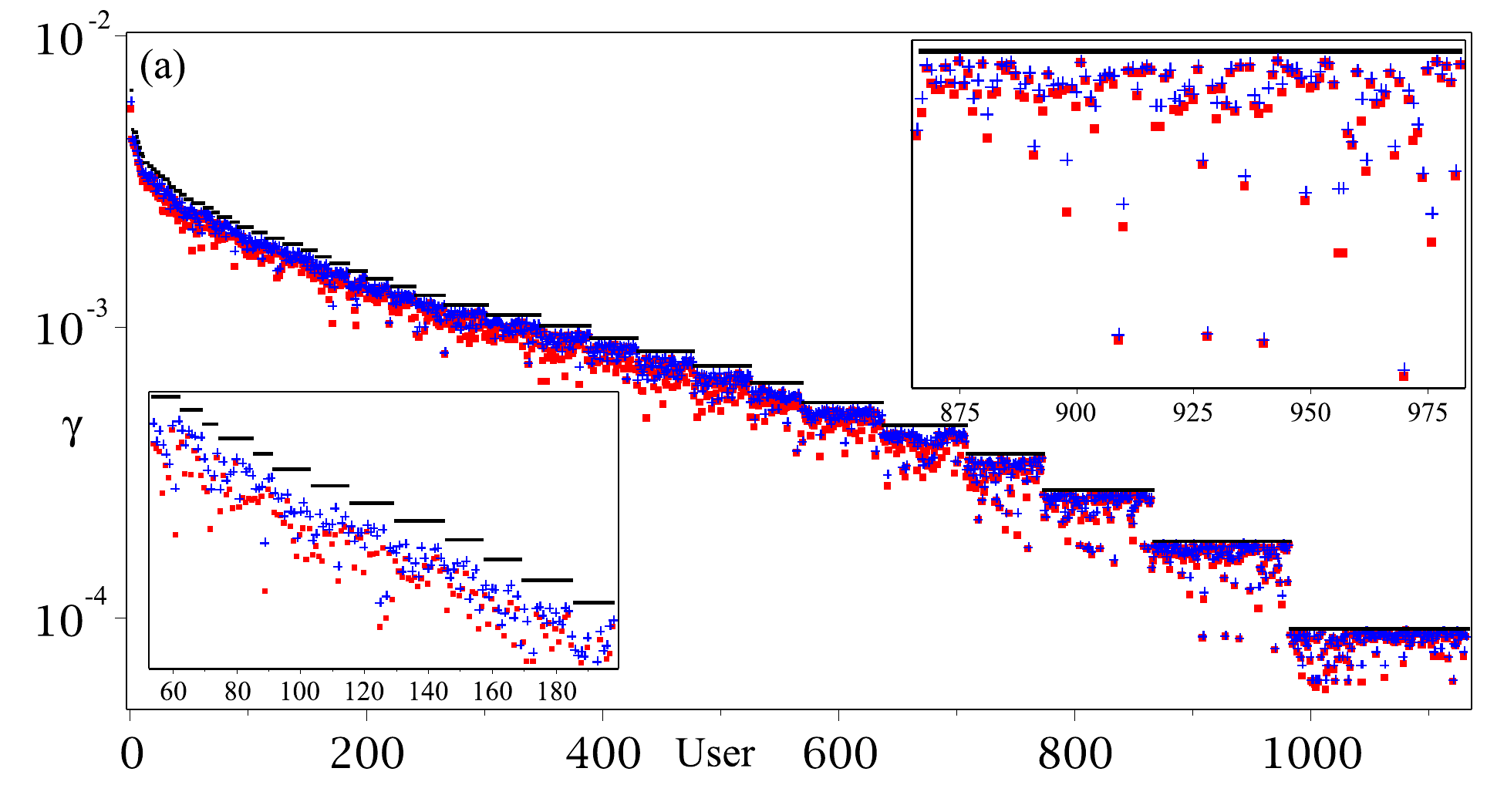}\\
 \includegraphics[width=8.5cm]{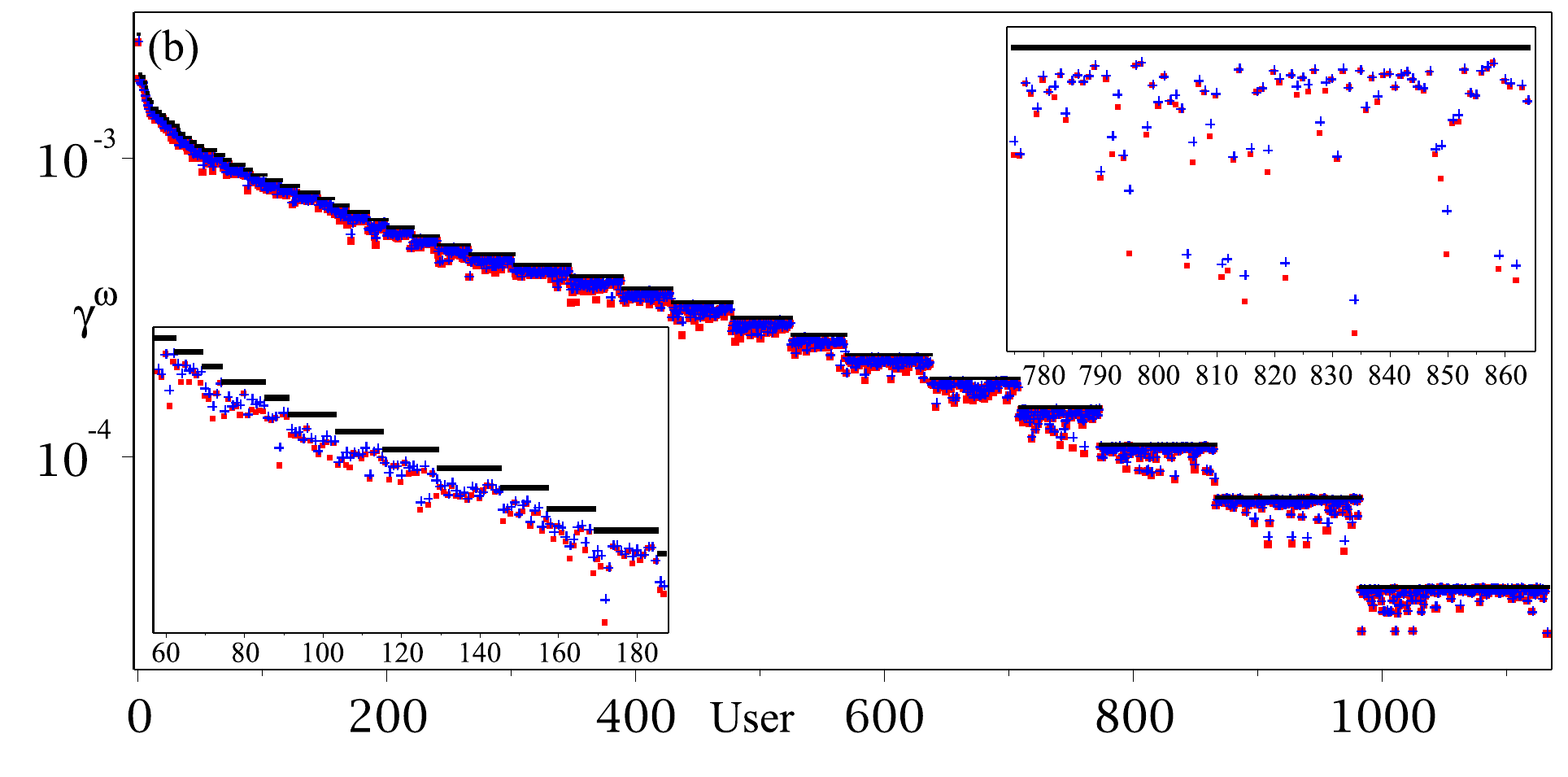}
\end{center}
\caption{(Colour online). Escape rate in the e-mail communication network. Illustrated here is the escape rate $\gamma^\omega$ (red
  squares), the equilibrium measure $\mu$ as in Eq.~(\ref{Eq:Gamma-first-order-form})
  (black line), and the approximation~$\hat{\gamma}$ from Eq.~(\ref{Eq:gamma_ap})
  (blue crosses) obtained if an individual e-mail user {\em fails}. (a) Fully open network
  $\omega=[0]^{\infty}$. (b) Temporal-varying failures with protocol $\omega=[01011]^{\infty}$. . The error -- as in Tab.~\ref{Table:AppAcc} -- between $(\gamma, \mu)$
  is $18.7$ and between $(\gamma,\hat{\gamma})$  is $4.52$. The {\em
    Spearman's}  rank  correlation coefficient $\rho_{A,B}$ between the ranks obtained
  from measures $A$ and $B$ are: 
$\rho_{\text{degree},\gamma^\omega}=$ 0.997,  $\rho_{\mu,\gamma^\omega}= 0.997$, and  $\rho_{\hat{\gamma},\gamma^\omega}=0.999$  for $\omega=[0]^\infty$;
 and
$\rho_{\text{degree},\gamma^\omega}=$ 0.781,  $\rho_{\mu,\gamma^\omega}= 0.993$, and  $\rho_{\hat{\gamma},\gamma^\omega}=1.000$ for $\omega=[01011]^{\infty}$. Finally,
comparing $\gamma^\omega$ of the two protocols we obtain $\rho=0.999$. Values reported to 3 s.f.
}
\label{fig:Emails1}
\end{figure}
The results shown in Fig.~\ref{fig:Airports1} and Fig.~\ref{fig:Emails1} confirm our conclusions obtained in the
random networks for failures in almost all nodes. In particular, we obtain that the
approximation~$\hat{\gamma}$ in Eq.~(\ref{Eq:gamma_ap}) is better than~$\mu$ in
Eq.~(\ref{Eq:Gamma-first-order-form}) and that (temporal) failures induce switches in the
rank of the nodes. As we saw in the scale-free networks previously, both approximations $\mu$ and $\hat{\gamma}$ systematically overestimate
$\gamma$.

On the other hand, a new
observation in the air-transportation network is the existence of sets of nodes which
deviate substantially from the approximations (the outliers in Fig.~\ref{fig:Airports1}):
the airports from $1$ to $73$ (hubs) and the set $S$ of five airports
$S=\{291,303,308,479,481\}$. Next we explain these deviations and their relation with the existence of communities in the network, i.e., nodes
that are strongly connected to each other and only weakly  connected to other
nodes\cite{NewmanBookNetworks}.  In the air transportation network the nodes $S$ are the
most interconnected ones, having a single weak connection (through node $13$) to the rest of the
network (a bottleneck).  The escape rate from $S$ to the outside is
$\gamma^*=0.00238$ (obtained, e.g., if we 
start in $S$ and consider the node $13$ to fail). If a node outside $S$ fails, two cases
have to be considered. For nodes with small degree, the escape rate from the large portion
of the network is $\gamma<\gamma^*$ and therefore it dominates the global escape rate and
agrees with the approximations. For nodes with high degree, the escape from 
the large portion of the network is fast, $\gamma> \gamma^*$, and hence the global rate
becomes $\gamma=\gamma^*$ 
because asymptotically the smallest escape rate dominates. This explains the appearance of
the plateau in the hubs of Fig.~\ref{fig:Airports1} at $\gamma\approx \gamma^*$. In this
case the approximations $\mu$ and $\hat{\gamma}$ describe closely the escape rate obtained
from the {\it second} largest eigenvalue of the network whose eigenvector is localized outside
$S$ (the eigenvector associated to $\gamma^*$ is always localized in $S$). Finally, if a node in $S$ fails, the escape rate is dominated by the small
probability to jump from the large network through the bottleneck (in node $13$) to $S$,
explaining the appearance of the same small $\gamma$ for all these nodes (bottom crosses
in  Fig.~\ref{fig:Airports1}ab). 

In order to understand further the effect that the topology of these real-world
  networks has, we repeat our analysis in  link randomised versions of the airport and
  e-mail networks.  In both cases we obtain that the escape rate is very well approximated
  by the zeroth order approximation~(\ref{eq.gamma0}) (Speraman's rank
  correlation of $1.000$  in the randomized airport network and $0.997$ in the e-mail
  network). This happens because the randomization destroys communities and short
  loops in the network, the sources of deviations from the zero-order approximation
  discussed above. Short loops are ubiquitous in real-world networks, in which improved
  results are obtained by our  approximation~(\ref{Eq:gamma_ap}).

\section{Conclusions}
\label{Sec:Out}

In summary, we have shown how to efficiently estimate the escape rate of random walkers in
networks in which specific nodes fail. Our new formula~(\ref{Eq:gamma_ap}) outperforms
estimations based on the  
degree or equilibrium measures, shows the importance of the degree and of short loops passing through the
failing nodes, and applies also to temporal dependent failures (stochastic or through a fixed
protocol).  The approximation is valid for small perturbations (i.e., degree of the
failing node much smaller than network size). In this limit, our results are expected to hold also in cases in which multiple nodes and edges fail.  

The escape rate can be viewed as a measure of node importance that quantifies how
sensitive   the network is to failures in this node (\emph{failure-centrality}). The results summarized above allow
for an estimation of this measure, which differs from the degree and other centrality
measures. For instance, the eigenvector and page-rank centrality coincide with the
  zero-th order approximation of the failure centrality, Eq.(\ref{eq.gamma0}), and are
  therefore different from the failure centrality already in this simple case. In networks
  with general degree distributions these measures differ even more significantly from the
  failure centrality, e.g., in the scale-free networks discussed in Table
  \ref{Table:AppAcc} we found a Spearman's rank correlation of 0.735 (to page rank) and
  0.748 (to eigenvector centrality). The failure-centrality measure depends also on the type of temporal failure the nodes
experience. We have shown how switches in the importance of nodes appear and can be
engineered depending on the interplay between the period of the temporal failure and the
length of the loops present in the network.

In real-world networks, the appearance of communities enhance the deviations between the
escape rate and naive estimations. At the same time, when hubs fail the escape of random
walks from the closest communities may dominate the global escape rate and lead to
deviations from our theory. The introduction of failures in nodes  can thus
be used to quickly detect the strongest communities of a network, in the same way that holes have been introduced in chaotic
dynamical systems  to visualize their underlying invariant manifolds~\cite{RMP}.

\emph{Acknowledgments} We thank Carl Dettmann and Oresits Georgiou for insightful
discussions.



\begin{thebibliography}{10}

\bibitem{NewmanBookNetworks}
M.~Newman.
\newblock {\em Networks: An Introduction}.
\newblock OUP Oxford, 2010.

\bibitem{Mon69}
Elliott~W. Montroll.
\newblock Random walks on lattices. iii. calculation of first passage times
  with application to exciton trapping on photosynthetic units.
\newblock {\em J. Math. Phys.}, 10(4):753--765, 1969.

\bibitem{TejEtAl09}
V.~Tejedor, O.~B\'enichou, and R.~Voituriez.
\newblock Global mean first-passage times of random walks on complex networks.
\newblock {\em Phys. Rev. E}, 80:065104, Dec 2009.

\bibitem{LinZhang13}
Yuan Lin and Zhongzhi Zhang.
\newblock Influence of trap location on the efficiency of trapping in
  dendrimers and regular hyperbranched polymers.
\newblock {\em J. Chem. Phys.}, 138(9):--, 2013.

\bibitem{AdEtAl01}
Lada~A. Adamic, Rajan~M. Lukose, Amit~R. Puniyani, and Bernardo~A. Huberman.
\newblock Search in power-law networks.
\newblock {\em Phys. Rev. E}, 64:046135, Sep 2001.

\bibitem{GuiEtAl02}
R.~Guimera, A.~Diaz-Guilera, F.~Vega-Redondo, A.~Cabrales, and A.~Arenas.
\newblock Optimal network topologies for local search with congestion.
\newblock {\em Phys. Rev. Lett.}, 89:248701, Nov 2002.

\bibitem{Holme03}
Petter Holme.
\newblock Congestion and centrality in traffic flow on complex networks.
\newblock {\em Advances in Complex Systems}, 06(02):163--176, 2003.

\bibitem{Noh04}
Jae~Dong Noh and Heiko Rieger.
\newblock Random walks on complex networks.
\newblock {\em Phys. Rev. Lett.}, 92:118701, Mar 2004.

\bibitem{PerraEtal12}
Nicola Perra, Andrea Baronchelli, Delia Mocanu, Bruno
  Gon\ifmmode~\mbox{\c{c}}\else \c{c}\fi{}alves, Romualdo Pastor-Satorras, and
  Alessandro Vespignani.
\newblock Random walks and search in time-varying networks.
\newblock {\em Phys. Rev. Lett.}, 109:238701, Dec 2012.

\bibitem{Newman2005}
M.E.~J. Newman.
\newblock {A measure of betweenness centrality based on random walks}.
\newblock {\em Soc. Networks}, 27(1):39--54, 2005.

\bibitem{RednerBookFirstPassage}
S.~Redner.
\newblock {\em A Guide to First-Passage Processes}.
\newblock Cambridge University Press, 2001.

\bibitem{Baronchelli2006}
Andrea Baronchelli and Vittorio Loreto.
\newblock Ring structures and mean first passage time in networks.
\newblock {\em Phys. Rev. E}, 73:026103, Feb 2006.

\bibitem{Lau2010}
H.~W. Lau and K.~Y. Szeto.
\newblock Asymptotic analysis of first passage time in complex networks.
\newblock {\em EPL}, 90(4):40005, 2010.

\bibitem{Afraimovich2010}
V~S Afraimovich and L.~A. Bunimovich.
\newblock {Which hole is leaking the most: a topological approach to study open
  systems}.
\newblock {\em Nonlinearity}, 23:643--656, 2010.

\bibitem{Geo12}
Orestis Georgiou, Carl~P. Dettmann, and Eduardo~G. Altmann.
\newblock Faster than expected escape for a class of fully chaotic maps.
\newblock {\em Chaos}, 22(4):--, 2012.

\bibitem{BunBook}
L.~Bunimovich and B.~Webb.
\newblock {\em Isospectral Transformations: A New Approach to Analyzing
  Multidimensional Systems and Networks}.
\newblock Springer Monographs in Mathematics. Springer New York, 2014.

\bibitem{BocEtal12}
Mohammud~Z. Bocus, Carl~P. Dettmann, Justin~P. Coon, and Mohammed~R. Rahman.
\newblock Keyhole and reflection effects in network connectivity analysis.
\newblock {\em CoRR}, abs/1211.6255, 2012.

\bibitem{Bun11}
Leonid~A. Bunimovich and Alex Yurchenko.
\newblock Where to place a hole to achieve a maximal escape rate.
\newblock {\em Israel J. Math.}, 182(1):229--252, 2011.

\bibitem{Cr13}
G~Cristadoro, G~Knight, and M~Degli~Esposti.
\newblock Follow the fugitive: an application of the method of images to open
  systems.
\newblock {\em J. Phys. A: Math. Theor.}, 46(27):272001, 2013.

\bibitem{RMP}
Eduardo~G. Altmann, Jefferson S.~E. Portela, and Tam\'as T\'el.
\newblock Leaking chaotic systems.
\newblock {\em Rev. Mod. Phys.}, 85:869--918, May 2013.

\bibitem{Bahsoun}
Wael Bahsoun and Sandro Vaienti.
\newblock Escape rates formulae and metastablilty for randomly perturbed maps.
\newblock {\em Nonlinearity}, 26(5):1415--1438, 2013-05-01T00:00:00.

\bibitem{Livorati}
Andr\'e L.~P. Livorati, Orestis Georgiou, Carl~P. Dettmann, and Edson~D.
  Leonel.
\newblock Escape through a time-dependent hole in the doubling map.
\newblock {\em Phys. Rev. E}, 89:052913, May 2014.

\bibitem{Martin2010}
O.~C. Martin and P.~\ifmmode~\check{S}\else \v{S}\fi{}ulc.
\newblock Return probabilities and hitting times of random walks on sparse
  erd\"os-r\'enyi graphs.
\newblock {\em Phys. Rev. E}, 81:031111, Mar 2010.

\bibitem{KeLi09}
Gerhard Keller and Carlangelo Liverani.
\newblock Rare events, escape rates and quasistationarity: Some exact formulae.
\newblock {\em Journal of Statistical Physics}, 135(3):519--534, 2009.

\bibitem{Hag08}
Aric~A. Hagberg, Daniel~A. Schult, and Pieter~J. Swart.
\newblock Exploring network structure, dynamics, and function using {NetworkX}.
\newblock In {\em Proceedings of the 7th Python in Science Conference
  (SciPy2008)}, pages 11--15, Pasadena, CA USA, August 2008.

\bibitem{Kim12}
H~Kim, C~I~Del Genio, K~E Bassler, and Z~Toroczkai.
\newblock Constructing and sampling directed graphs with given degree
  sequences.
\newblock {\em New J. Phys.}, 14(2):023012, 2012.

\bibitem{Bodai2013}
Tam\'as B\'odai, Eduardo~G. Altmann, and Antonio Endler.
\newblock Stochastic perturbations in open chaotic systems: Random versus noisy
  maps.
\newblock {\em Phys. Rev. E}, 87:042902, Apr 2013.

\bibitem{Main92}
Ronnie Mainieri.
\newblock Cycle expansion for the lyapunov exponent of a product of random
  matrices.
\newblock {\em Chaos}, 2(1):91--97, 1992.

\bibitem{VulBook93}
A.~Crisanti, G.~Paladin, and A.~Vulpiani.
\newblock {\em Products of Random Matrices in Statistical Physics}.
\newblock Springer Series in Solid-State Sciences. Springer-Verlag, 1993.

\bibitem{Colizza2007}
Vittoria Colizza, Romualdo~P. Satorras, and Alessandro Vespignani.
\newblock {Reaction-diffusion processes and metapopulation models in
  heterogeneous networks}.
\newblock {\em Nat Phys}, 3(4):276--282, April 2007.

\bibitem{Gu03}
Roger GuimerÃ, Leon Danon, Albert DÃ­az-Guilera, Francesc Giralt, and Alex
  Arenas.
\newblock Self-similar community structure in a network of human interactions.
\newblock {\em Phys. Rev. E}, 68(6):065103, 2003.

\bibitem{New08}
Mark Newman.
\newblock The physics of networks.
\newblock {\em Phys. Today}, 61(11):33--38, 2008.

\bibitem{ResEtal06}
Juan~G. Restrepo, Edward Ott, and Brian~R. Hunt.
\newblock Characterizing the dynamical importance of network nodes and links.
\newblock {\em Phys. Rev. Lett.}, 97:094102, Sep 2006.

\end{thebibliography}

\end{document}